\documentclass[preprint,1p,times]{elsarticle}

\usepackage{lineno,hyperref}
\modulolinenumbers[5]

\usepackage[usenames]{color}
\usepackage{multirow}

\usepackage{marginnote}
\usepackage{lipsum}                     
\usepackage[pdftex,dvipsnames]{xcolor}  
\usepackage{amsmath}
\usepackage[colorinlistoftodos,prependcaption,textsize=scriptsize]{todonotes}
\usepackage{xargs} 

\usepackage{natbib}

\usepackage{orcidlink}

\journal{Nucl. Instr. and Meth. Phys. Res. Sect. A}

\begin{document}

\begin{frontmatter}

\title{Common Iron Passive Magnetic Shielding for Scintillator Array Photomultiplier Tubes}


\author[1,2]{T.~Atovullaev}
\author[1,3]{S.~Cherepanov}
\author[1]{A.~Atovullaeva}
\author[4,5]{J.~Kahlbow}
\author[1]{{M.~Patsyuk}
\corref{corr1}}
\ead{mpatsyuk@jinr.ru}
\author[4]{O.~Hen}
\author[5]{G.~Johansson}
\author[1]{I.~Kruglova}
\author[5]{B.~Meirovich}
\author[1]{S.~Nepochatykh}
\author[5]{E.~Piasetzky}
\author[1]{S.~Piyadin}
\author[1]{A.~Salamatin}
\author[1]{S.~Sedykh} 
\author[5]{S.~Segev}
\author[6]{Y.~Zhang}

\cortext[corr1]{Corresponding author}

\address[1]{Joint Institute for Nuclear Research, Dubna, Russia}
\address[2]{Institute of Nuclear Physics, Ministry of Energy of the Republic of Kazakhstan, Almaty, Kazakhstan}
\address[3]{Lomonosov Moscow State University, Moscow, Russia}
\address[4]{Massachusetts Institute of Technology, Cambridge, USA}
\address[5]{School of physics and Astronomy Tel-Aviv University, Tel-Aviv, Israel}
\address[6]{Tsinghua University, Beijing, China}

\begin{abstract}

In accelerator-based experiments, detector arrays constructed from organic plastic scintillators are widely used for measurements of particle timing, hit position, and energy loss. These detectors are often operated in proximity to magnetic spectrometers, where the associated photomultiplier tubes (PMTs) are exposed to fringe magnetic fields. The standard approach to mitigate magnetic interference involves individually shielding each PMT. In this work, we present a combined passive shielding solution consisting of a common iron enclosure for a row of PMTs, supplemented with individual mu-metal cylinders. This configuration was deployed during a 2022 experiment at JINR. Performance evaluations show that the proposed shielding effectively preserves PMT gain and timing resolution, demonstrating its viability for future applications in similar experimental environments.

\end{abstract}

\begin{keyword}
passive magnetic shielding of PMTs; TOF; scintillator
\end{keyword}

\end{frontmatter}



\section{Introduction}
\label{sec:intro}

Modules of plastic scintillating detectors are commonly used for timing measurements in particle physics experiments~\cite{hades_tof,clas12_tof,r3b_tof}. Arrays of such detectors can support momentum determination for known particles, enable particle identification for particles with known momenta, and provide timing signals for trigger systems. To achieve broad solid angle coverage while maintaining high spatial resolution, these arrays typically consist of many long, narrow plastic scintillator bars. Each scintillator is usually instrumented with two photomultiplier tubes (PMTs), one at each end.
When scintillator arrays are operated near external magnetic fields -- such as those from spectrometer dipole magnets -- PMTs are susceptible to performance degradation. The fringe magnetic field can deflect electron trajectories inside the PMT, resulting in reduced gain or even complete signal loss.
Silicon photomultipliers (SiPMs) mitigate this issue due to their short carrier drift paths, however, they suffer from a relatively small active collection area, which limits their efficiency in the discussed application.
To mitigate these effects, PMTs are typically shielded using high permeability materials, such as mu-metal or soft iron, often enclosed in steel cylindrical housings. Compensation coils can also be employed, particularly for axial magnetic fields, but their effectiveness is limited in complex, high-gradient fringe fields. Dynamic magnetic shielding~\cite{dynamic_shield} offers another alternative but requires active field monitoring and introduces additional complexity in fabrication and operation.
As a simpler and more scalable alternative, we investigate a passive shielding approach that uses a common iron enclosure for an entire row of PMTs, with each PMT further enclosed in its own mu-metal cylinder.
In this work, we report on the successful implementation and performance of this shielding method in the SRC experiment at BM@N in 2022 at Joint Institute for Nuclear Research (JINR, Russia), where the dipole analyzing magnet operated at 8\,kG and produced fringe fields of approximately 25\,G in the PMT region.

\section{Experimental conditions}
\label{sec:exp}

\begin{figure}[!htb]
    \centering
    \includegraphics[width=0.8\linewidth]{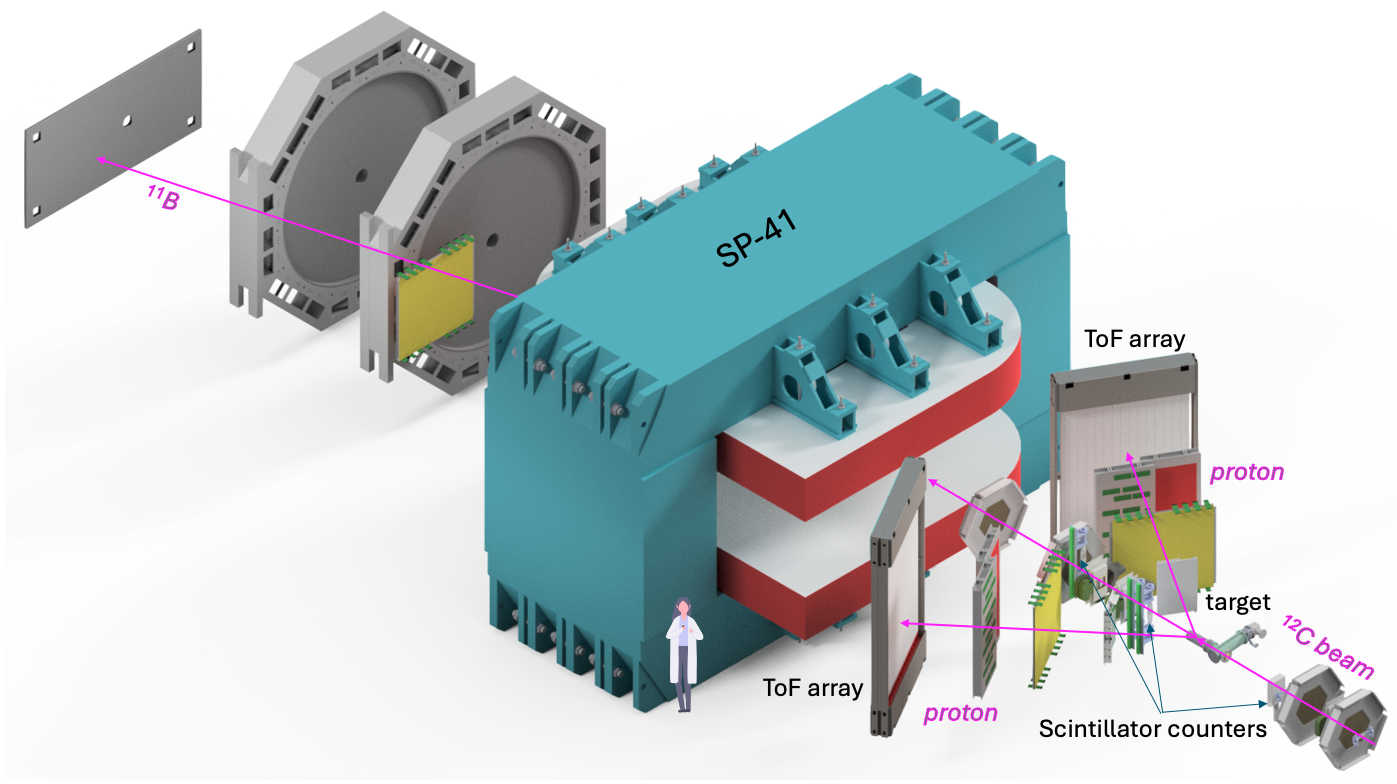}
    \caption{The experimental setup for the hard quasi-elastic $^{12}$C(p,2p)$^{11}$B measurement at BM@N in 2022.}
    \label{fig:setup}
\end{figure}

A common passive magnetic shielding system was developed for the two time-of-flight (ToF) scintillator arrays used in the two-arm spectrometer, which was a part of the special configuration of the BM@N setup~\cite{bmn_setup} for the SRC measurement in 2022. The SRC experiment was aimed to investigate hard quasi-elastic free $^{12}$C(p,2p)$^{11}$B scattering in inverse kinematics, using a 3.7\,GeV/c/u $^{12}$C beam incident on a cryogenic liquid hydrogen target~\cite{target}. The primary goal was to explore nuclear structure and single-particle dynamics.
An overview of the experimental setup is presented in Fig.~\ref{fig:setup}. The outgoing protons from the $^{12}$C(p,2p)$^{11}$B reaction were detected at laboratory angles of $\pm$30$^{\circ}$ relative to the beam axis using a dedicated two-arm spectrometer equipped with coordinate detectors and ToF arrays. Beam scintillator counters, positioned upstream and downstream of the target, were used to determine the event start time and to measure the total charge per event. Precise timing calibration of the scintillator counters was achieved using a dedicated laser calibration system~\cite{laser_calibration}, allowing alignment and validation of detector timing independent of beam operation.

Events of interest were selected based on coincidence signals from both ToF arrays, along with identification of an incoming $^{12}$C nucleus striking the target and a charged fragment emerging from the interaction.

\begin{figure}[!htb]
    \centering
    a)
    \includegraphics[width=0.35\linewidth]{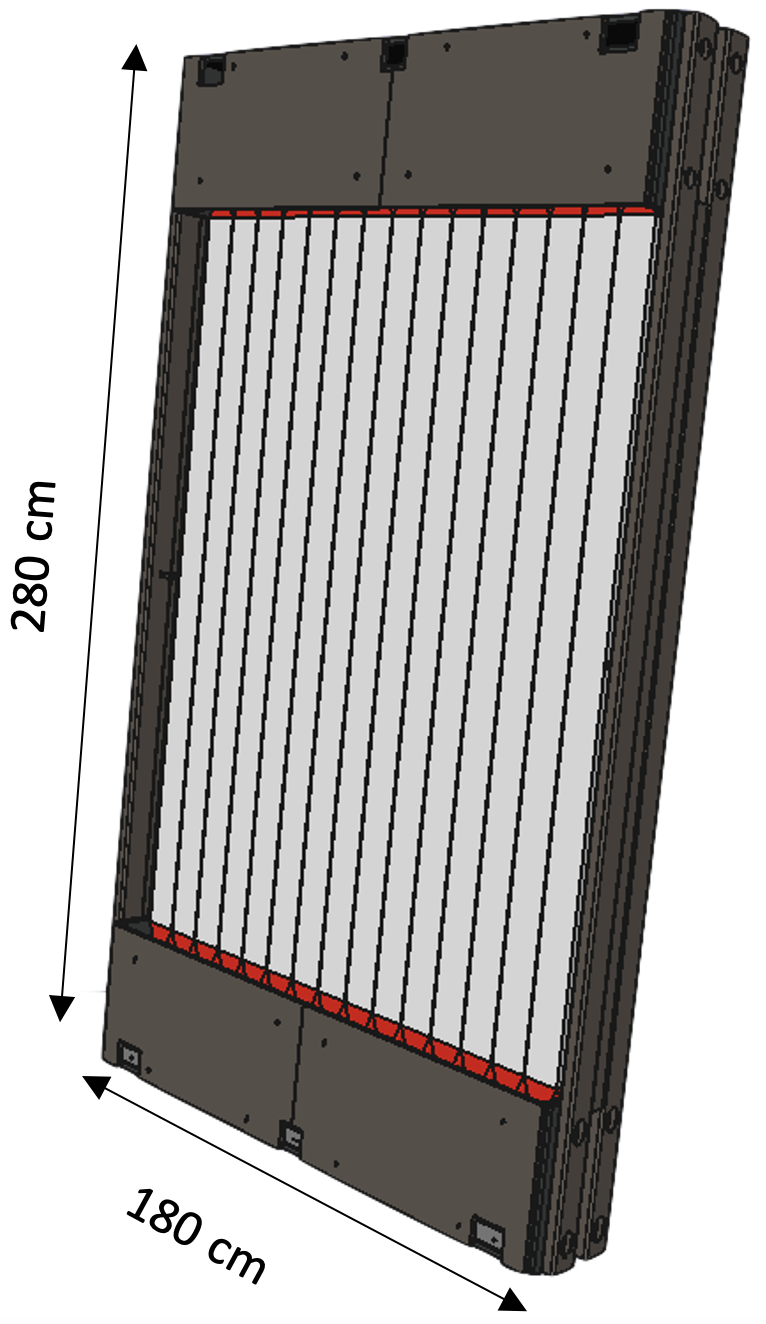} \hspace{1cm}
    b)
    \includegraphics[width=0.35\linewidth]{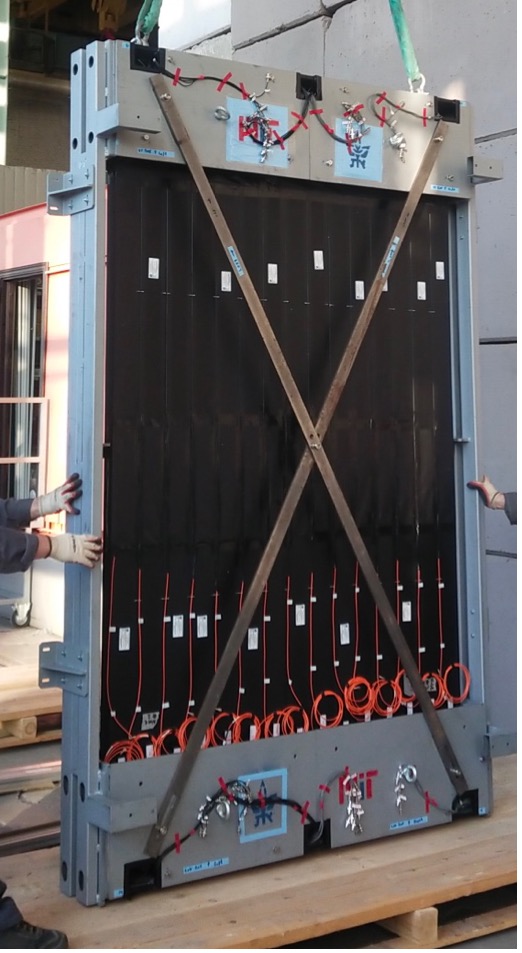}
    \caption{(a) Technical drawing and (b) photograph of a ToF array containing 15 scintillator counters. The PMTs are located behind the steel covers at the array’s top and bottom ends.}
    \label{fig:tof}
\end{figure}

Each ToF array (Fig.~\ref{fig:tof}) consisted of 15 vertically aligned plastic scintillator bars (EJ200) with dimensions of 10\,cm (width) × 6\,cm (thickness) × 200\,cm (length). Each scintillator bar was optically coupled via plastic light guides to a pair of 2-inch Hamamatsu R13435 PMTs~\cite{pmt_R13435}, one at each end. The PMTs were fixed behind a protective steel cover located at the top and bottom of each array.

The spectrometer’s warm dipole analyzing magnet (SP-41), powered by direct current through copper coils, has coil dimensions of 5.2\,m (beam direction) × 4.3\,m (transverse) × 0.6\,m (vertical), enclosed within a steel yoke measuring 2.5\,m × 7.0\,m × 4.5\,m. The vertical aperture between the magnet poles is 1.1\,m, with pole dimensions of 2.5\,m × 1.4\,m. The magnet operated at a current of 1650\,A, producing a central magnetic field of approximately 8\,kG.
The ToF arrays were positioned roughly 0.5\,m from the dipole magnet, where simulations using Ansoft Maxwell 15.0~\cite{ansoft} estimated the fringe magnetic field to be approximately 25\,G~\cite{sim}. For R13435 PMTs, even a transverse magnetic field of 10 G can result in a gain reduction of nearly an order of magnitude~\cite{sim}. Thus, effective magnetic shielding was essential to ensure stable PMT operation.

The following sections describe the shielding system’s design, magnetic field measurements, and performance evaluation through comparative measurements taken with and without the magnetic field. These results demonstrate the efficacy of the implemented passive shielding solution.

\section{Design and construction of the shielding boxes}
\label{sec:design}

To mitigate the effects of the fringe magnetic field, common iron shielding boxes were designed to reduce the magnetic field at the PMT locations to levels where individual mu-metal shielding becomes effective. The shielding design was guided by simulations described in~\cite{sim}, which modeled monolithic steel enclosures with the wall thicknesses of 5\,mm, 10\,mm, and 20\,mm, and magnetic permeabilities of 2500, 5000, and 10000.

The simulations indicated that the internal magnetic field showed relatively weak dependence on the permeability value, while increasing the wall thickness significantly improved the shielding performance. Based on this, the final design employed 14\,mm thick walls made of Steel-15, a material with appropriate magnetic properties and mechanical robustness.

\begin{figure}[!htp]
    \centering
    \includegraphics[width=0.55\linewidth]{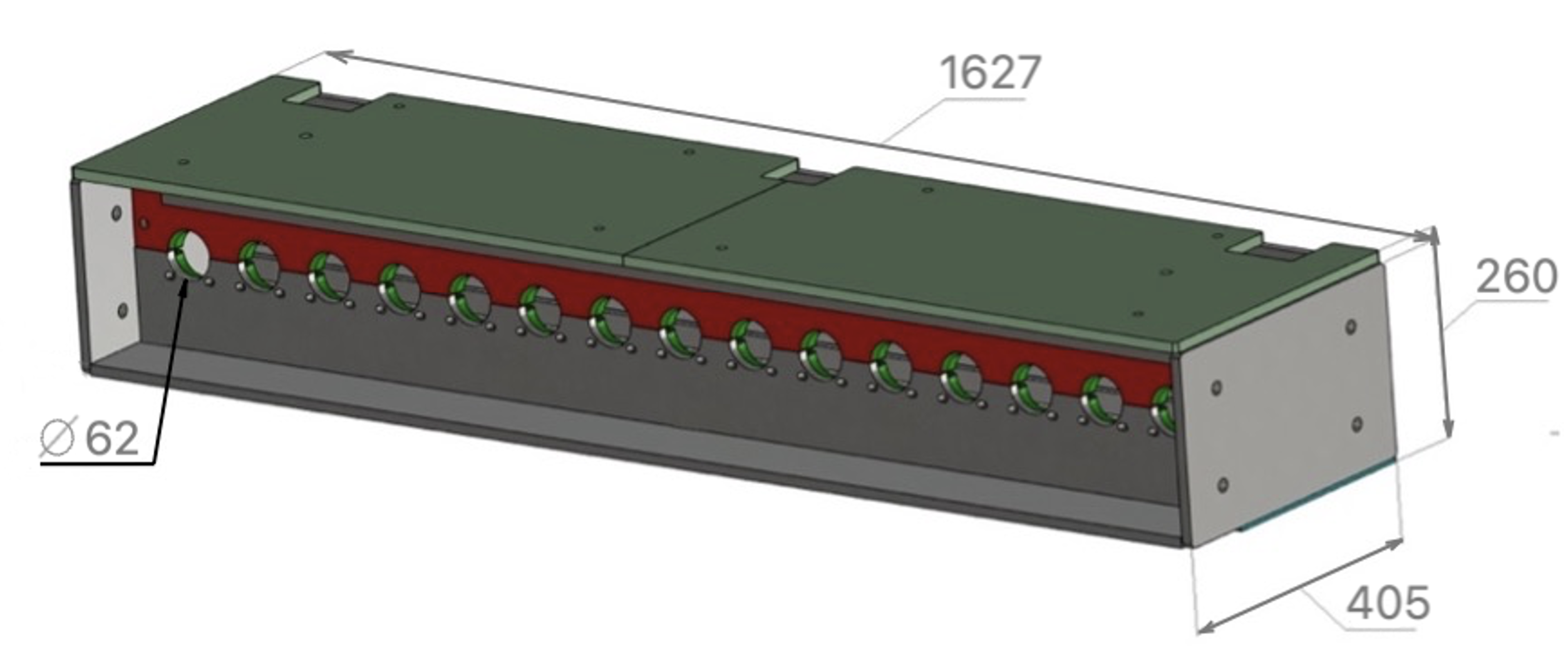}
    \caption{Technical drawing of the common passive magnetic shielding box for a row of fifteen PMTs. Dimensions are shown in mm.}
    \label{fig:box}
\end{figure}

Each shielding box enclosed an entire row of 15\,PMTs located at the top and bottom of the ToF arrays. The structure was assembled from painted steel sheets and securely mounted to the detector frame using screws, as illustrated in Fig.~\ref{fig:box}. The front panel of each box included three rectangular cutouts to allow for the routing of signal and high-voltage (HV) cables.

\begin{figure}[!htb]
  \begin{minipage}[t]{.49\textwidth}
    \includegraphics[width=\textwidth]{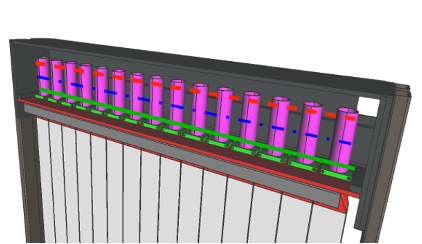}
  \end{minipage}
  \hfill
  \begin{minipage}[t]{.49\textwidth}
    \includegraphics[width=\textwidth]{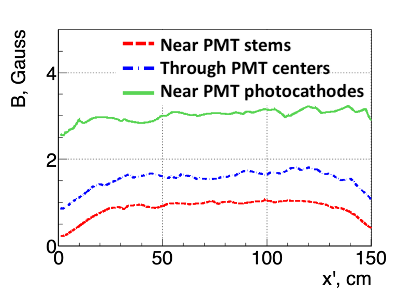}
  \end{minipage}
   \caption{Simulated magnetic field strength inside the shielding box along three horizontal lines running through the PMT axes (mu-metal cylinders not included in the simulation): near the PMT stems (dashed red), through the PMT centers (dash-dotted blue), and near the PMT photocathodes (solid green).}
    \label{fig:pmts}
\end{figure}

Simulations showed that the magnetic field at the PMT positions inside the shielding box was reduced to below 3.5\,G, as illustrated in Fig.~\ref{fig:pmts}. However, to further suppress residual magnetic fields -- particularly the longitudinal component that PMTs are most sensitive to -- additional individual shielding was applied around each PMT.

Each PMT was enclosed in a mu-metal cylinder with an inner diameter of 57\,mm and a length of 205\,mm, constructed from two layers of 1.57\,mm thick spot-welded Ad-Mu-80 alloy, a high-permeability material~\cite{mu-metal}. The thickness of the mu-metal tubes was based on previous studies~\cite{band}, which demonstrated effective suppression of fringe magnetic fields in similar experimental conditions.
\section{Results: PMT performance with and without magnetic field}
\label{sec:results}

The response of PMTs was evaluated for signal amplitude stability and timing resolution both in the presence and absence of the magnetic field. Data were collected using a $^{12}$C beam incident on a hydrogen target, with final-state fragments detected in the forward (beam) direction. Analysis focused on events where detected fragments carried charges less than four times the proton charge, yielding well-defined pulses in the ToF counters. Additional calibration data were obtained by injecting laser pulses into the center of the scintillator bars.

\begin{figure}[!htb]
    \centering
    a)
    \includegraphics[width=0.3\linewidth]{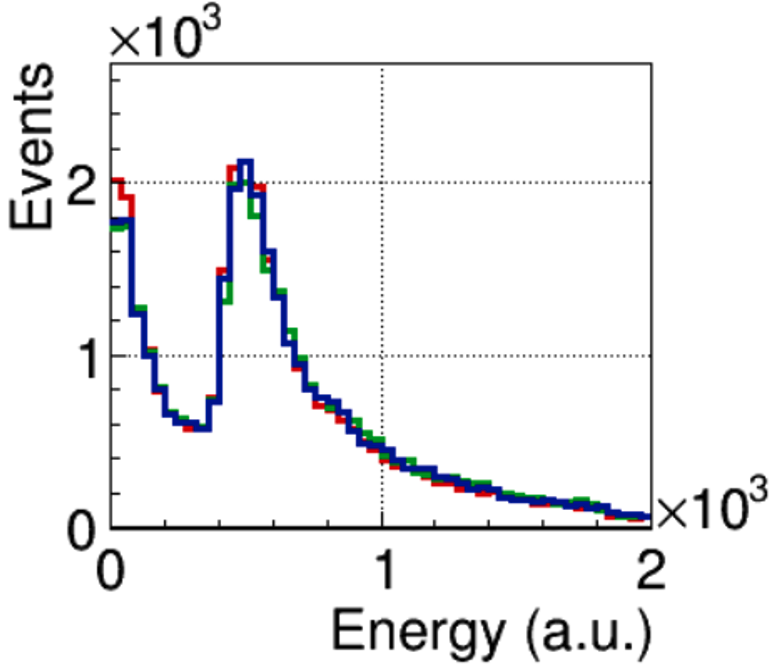}
    b)
    \includegraphics[width=0.3\linewidth]{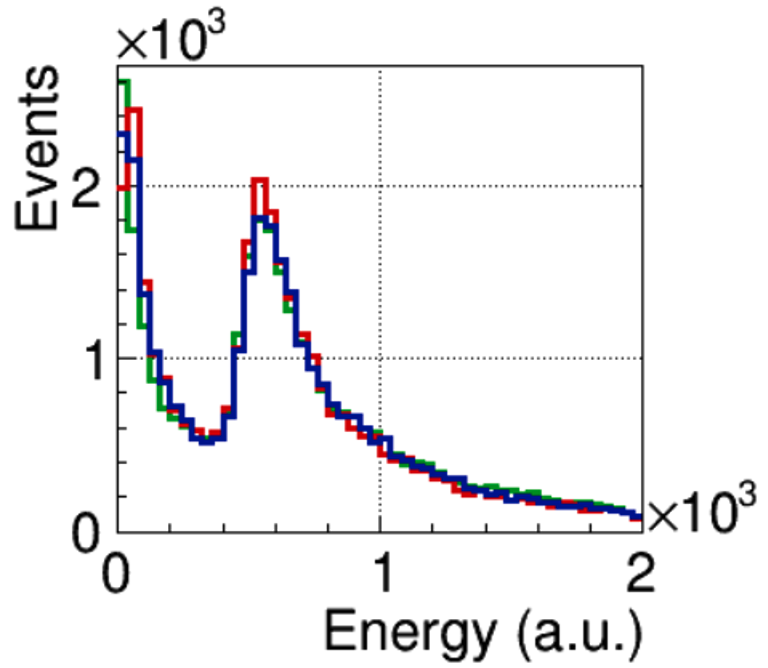}
    \caption{Geometric mean of signal amplitudes from upper and lower PMTs across all 15 scintillator bars for the left (a) and right (b) spectrometer arms. Blue corresponds to magnetic field off; red and green correspond to measurements with the field on, taken before and after the no-field period, respectively.}
    \label{fig:res1}
\end{figure}

Measurements were performed with the large dipole magnet alternately switched on and off, enabling direct comparison of PMT response with and without magnetic field. 
Figure~\ref{fig:res1} presents the geometric mean of signal amplitudes from the upper and lower PMTs across all 15 scintillator bars in both the left and right detector arms. 
The data include three configurations: magnetic field off (blue), and field on before (red) and after (green) the no-field period. Each distribution contains an equal number of events. The position and shape of the Minimum Ionizing Particle (MIP) peak, observed around 500–600\,a.u., remains consistent across all configurations.
The background distribution peaking at low amplitudes and spreading up to high energy values is caused by particles having different path lengths inside the scintillator bar.
The consistency between three distributions indicates stable PMT gain and validates the efficiency of the passive magnetic shielding.

\begin{figure}
    \centering
    a)
    \includegraphics[clip, width=0.45\linewidth]{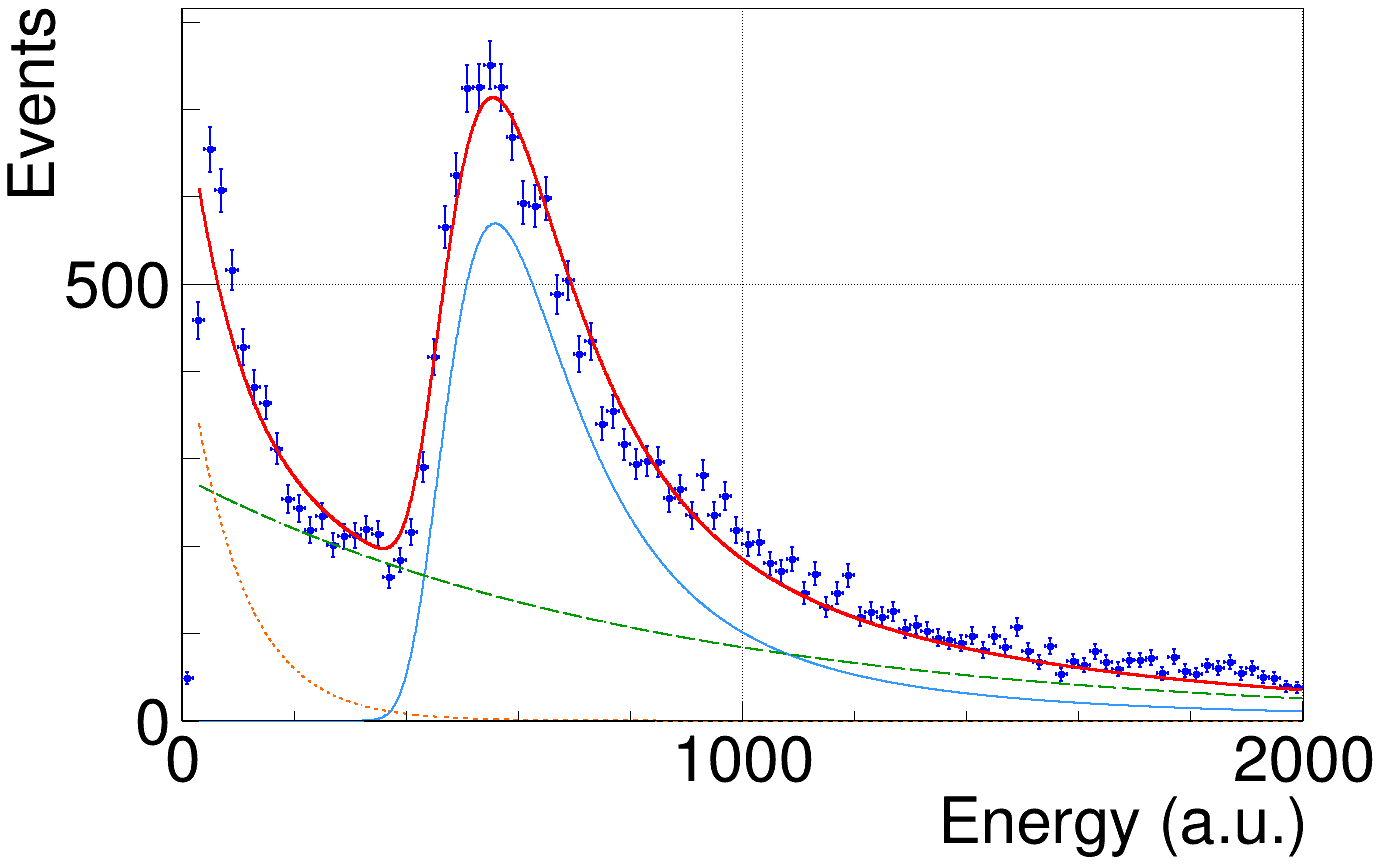}
    b)
    \includegraphics[clip, width=0.45\linewidth]{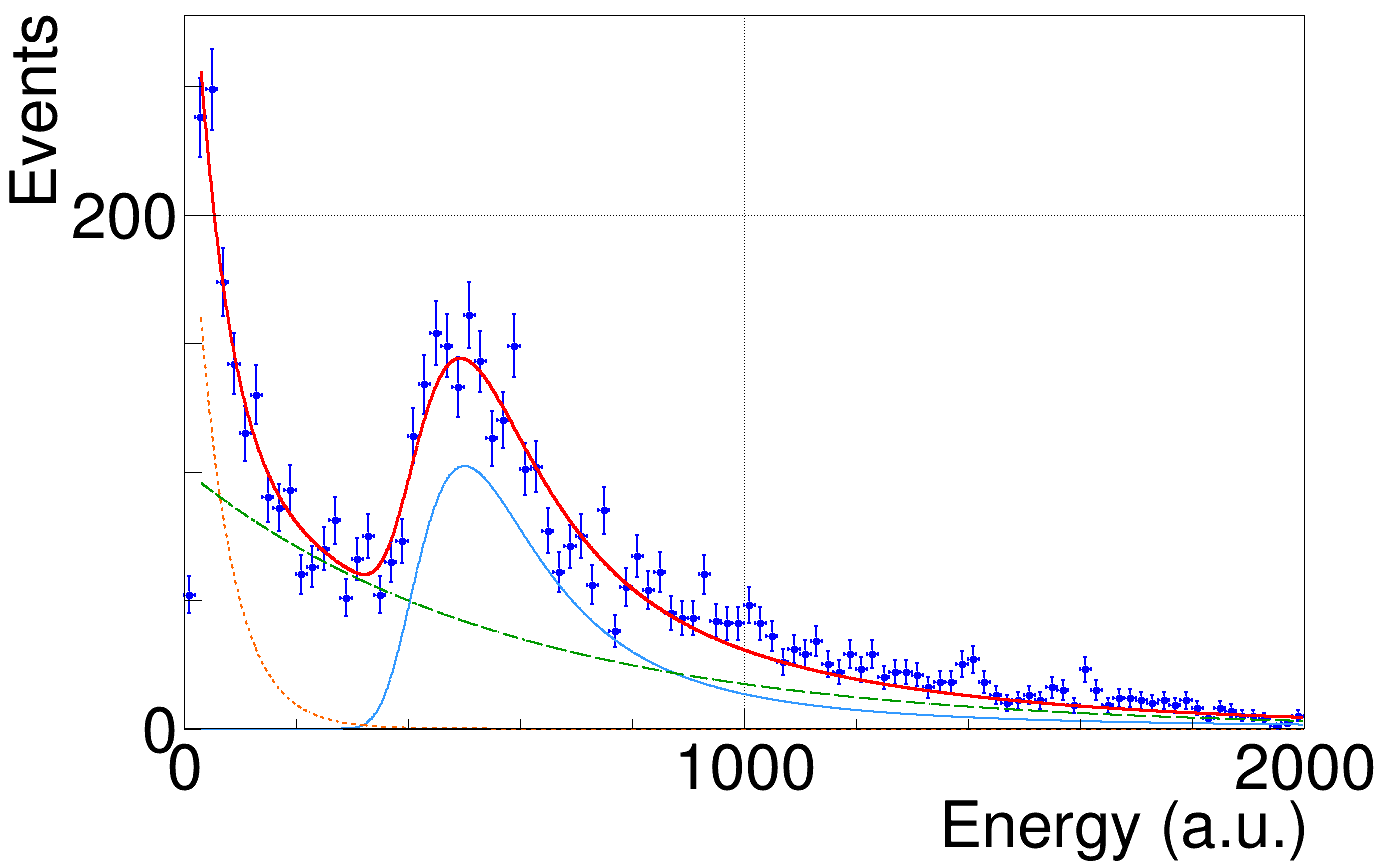}
    \caption{PMT amplitude distributions with fits to a combination of a Landau distribution and two exponents. See text for details.}
    \label{fig:chi2}
\end{figure}

\begin{figure}
    \centering
    a)
    \includegraphics[clip, width=0.447\linewidth]{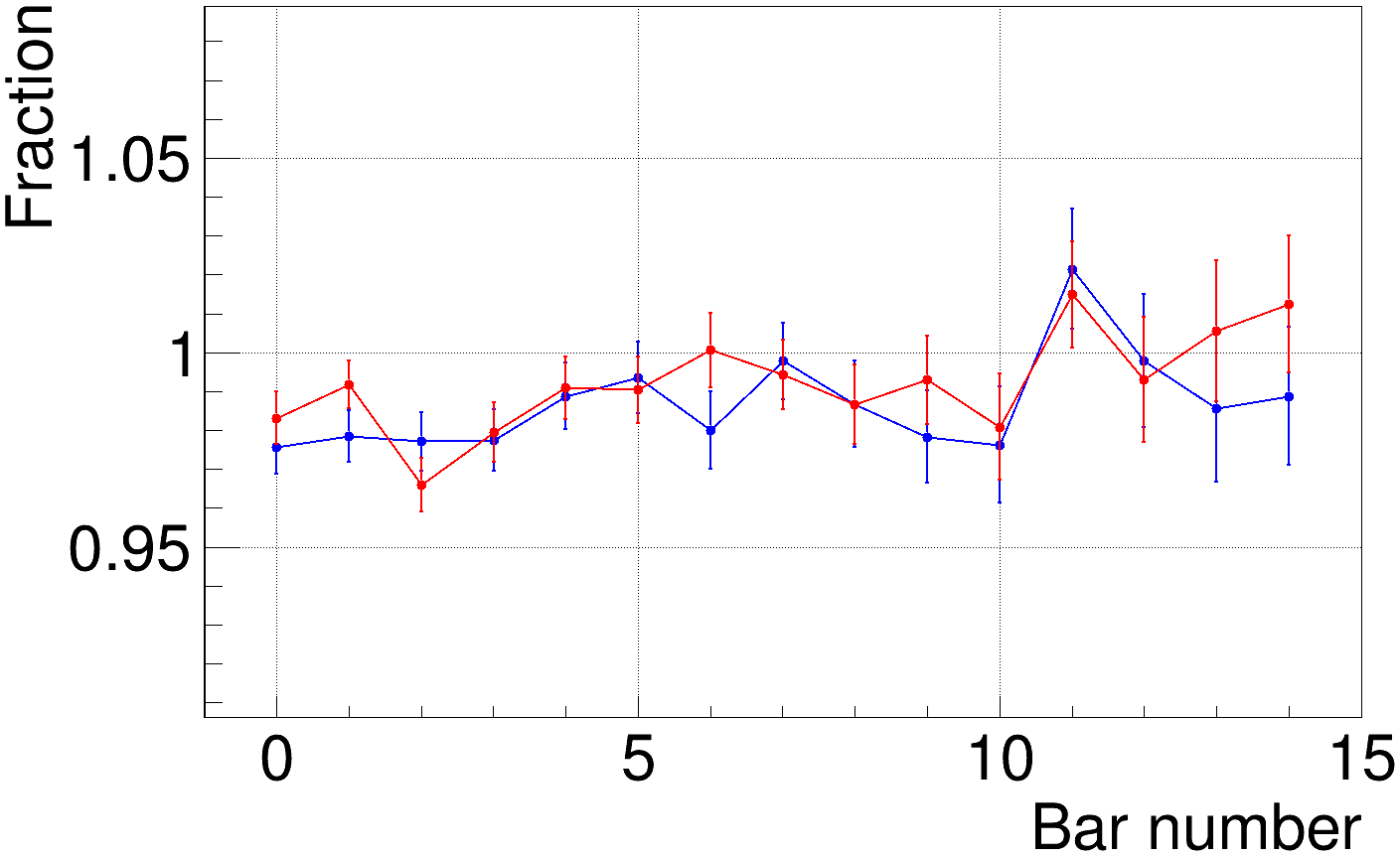}
    b)
    \includegraphics[clip, width=0.45\linewidth]{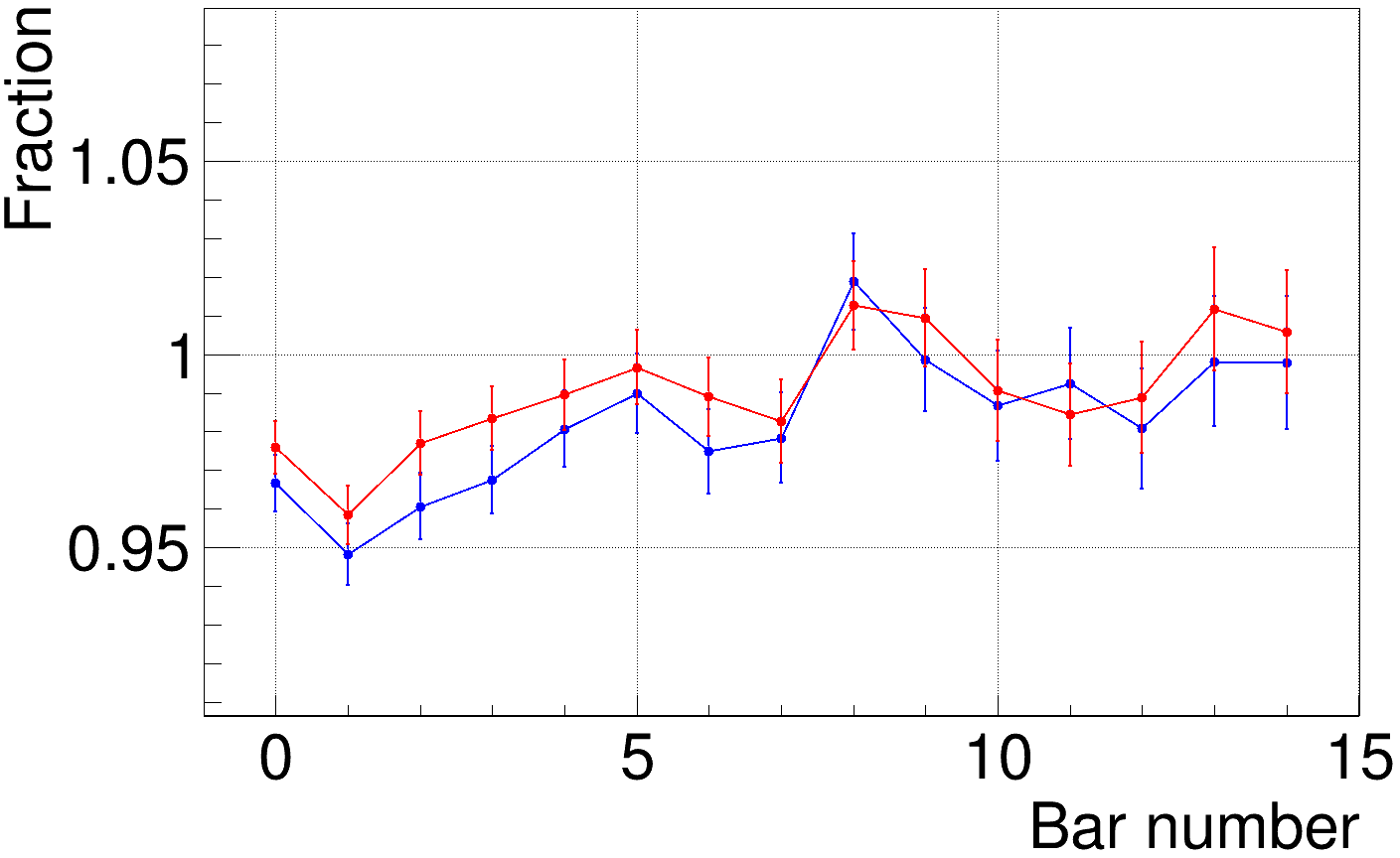}
    \caption{The ratios of the most probable values and mean values for the amplitude spectrum for physics data as a function of PMT number for one of the PMT The ratios of the most probable values and mean values for the amplitude spectrum for physics data as a function of PMT number for one of the PMT rows.}
    \label{fig:tim}
\end{figure}

To further probe amplitude stability, the amplitude spectra from each PMT recorded for physics data were fitted with a combination of a Landau distribution and two exponents, see Fig.~\ref{fig:chi2}. 
The stability of the mean obtained from the histogram and the most probable values obtained as an approximating function, as a parameter of the Landau distribution, of the amplitude spectra were investigated for the three configurations using the ratios between the measurements with and without magnetic field.
The resulting distributions are shown in Fig.~\ref{fig:tim} ((a) for the most probable value, (b) for the mean) : the ratio for the measurement before/after the magnetic field was turned off is shown in blue/red. The two ratios are close to each other within error bars, which limits the influence of the magnetic field to the amplitude to be at the order of 4\%. The maximal influence is observed for the PMTs closest to the magnet, and for the other PMTs the amplitude change is negligible.

Timing performance under varying magnetic field conditions was evaluated using laser-triggered data. A time-walk correction was applied individually to each PMT. 
The individual time resolution of each PMT was extracted from time-difference distributions between pairs of PMTs selected from different bars in order to obtain statistical independence of their response to the laser pulses. These distributions were fitted with Gaussian functions, and the resulting widths $\sigma_{i,j}$ were calculated assuming independent contribution from the two PMTs, 
i.e. $\sigma_{i,j}^2 = \sigma_i^2 + \sigma_j^2$. If pairs from a set of three PMTs are selected for such time-difference distributions, the resulting system of three 
equations allows determination of all three contributing individual time resolutions.
All possible combinations for selection of sets of three PMTs were considered
in a systematic way, and the final individual time resolutions were obtained by averaging the extracted $\sigma$ values over all combinations.
The peak positions remained stable, the widths were within 60–100\,ps, and no significant deviations were observed for the three magnetic field configurations, confirming that the magnetic field has a negligible effect on PMT timing performance.

\section{Conclusions}
\label{sec:conc}

We have developed, fabricated, and successfully implemented a common passive iron magnetic shielding system for an array of PMTs, complemented by individual mu-metal cylinders around each PMT. This shielding configuration was deployed during the second SRC experiment at BM@N (JINR, Russia) in 2022 to protect PMTs from the external fringe magnetic field of approximately 25\,G.

Our results demonstrate that this combined shielding approach effectively preserves PMT performance, enabling stable operation without gain degradation or deterioration in time resolution. The solution provides a practical, cost-effective, and scalable method for magnetic field mitigation in environments with complex fringe fields, and is therefore well suited for future applications in high-energy and nuclear physics experiments involving large PMT arrays operating near strong magnetic spectrometers.



\section*{Acknowledgements}

We extend our sincere gratitude to the engineering team of the experimental hall 205 at JINR VBLHEP, led by S. Yu. Anisimov, for their crucial role in the successful setup of the detector and for their steadfast support, which were essential to the progress of our work.




\end{document}